# Structure analysis of the Ga-stabilized GaAs(001)-c(8x2) surface at high temperatures

Akihiro Ohtake*
*National Institute for Materials Science (NIMS), Tsukuba 305-0047, Japan*
*and Joint Research Center for Atom Technology (JRCAT), Tsukuba 305-0046, Japan*

Shiro Tsukamoto, Markus Pristovsek, and Nobuyuki Koguchi
*National Institute for Materials Science (NIMS), Tsukuba 305-0047, Japan*

Masashi Ozeki
*Joint Research Center for Atom Technology (JRCAT), Tsukuba 305-0046, Japan*
*and Department of Electrical and Electronic Engineering, Faculty of Engineering, Miyazaki University, Miyazaki 889-2192, Japan.*

Structure of the Ga-stabilized GaAs(001)-c(8x2) surface has been studied using rocking-curve analysis of reflection high-energy electron diffraction (RHEED). The c(8x2) structure emerges at temperatures higher than 600°C, but is unstable with respect to the change to the (2x6)/(3x6) structure at lower temperatures. Our RHEED rocking-curve analysis at high temperatures revealed that the c(8x2) surface has the structure which is basically the same as that recently proposed by Kumpf et al. [Phys. Rev. Lett. **86**, 3586 (2001)]. We found that the surface atomic configurations are locally fluctuated at high temperatures without disturbing the c(8x2) periodicity.

The (001) surface of compound semiconductors, such as GaAs and InAs, shows a variety of reconstructions depending on the processing condition and the resultant surface composition. The As-stabilized (2x4) surface of GaAs(001) has been most extensively studied, and is widely accepted to have the two As-dimer model (so-called É¿2 model).[1] On the other hand, despite considerable efforts, no well-established model was proposed for the atomic structures of the Ga-stabilized surface of c(8x2). The three Ga-dimer model [Fig. 1(a)] has been supported by the I-V curve analysis of low-energy electron diffraction (LEED),[2] but was found to be energetically unstable.[3] On the other hand, although it appears that the two-dimer model [Fig. 1(b)][4] and the top As-dimer model [Fig. 1(c)][5] are consistent with scanning tunneling microscopy (STM) images, both models failed in LEED test.[2]

Recent studies based on first-principles calculations have altogether changed such a situation:[6] Lee, Moritz, and Sheffler proposed the É$f$ model [Fig. 1(d)] which explains well LEED and STM data,[6] and is also supported by x-ray diffraction (XRD) analysis.[7] A similar structure model shown in Fig. 1(e) has been proposed by an independent study on the basis of the XRD analysis using direct methods.[8,9] The two models shown in Figs. 1(d) and 1(e) have basically the same atomic structures, but differ in the presence of Ga adatoms and in the partial absence of surface Ga-dimers in the latter model [Fig. 1(e)].[8,9]

Both models in Figs. 1(d) and 1(e) explains well experimental data obtained at room temperature.[6-9] However, as we will show later in this paper, the c(8x2) structure is stable only at temperatures higher than 600°C, but reversibly changes to the (2x6)/(3x6) structures as the temperature is decreased. Thus, in order to study the actual atomic structure of the c(8x2) surface without considering possible coexistence of other phases, structure analysis at high temperatures is indispensable.

This paper reports the surface structure analysis of the GaAs(001)-c(8x2) at high temperatures (> 600°C). Rocking-curve analysis of reflection high-energy electron diffraction (RHEED) based on dynamical diffraction theory[10-13] has been used for this purpose. The results show that the GaAs(001)-c(8x2) has the structure model which is basically the same as that proposed by Kumpf et al.[8,9] However, the formation of surface Ga-dimers is not necessarily favorable at high temperatures.

The experiments were performed in a dual-chamber molecular-beam epitaxy system which is equipped with an x-ray photoelectron spectroscopy and a STM for on-line surface characterization. Cleaned GaAs(001)-(2x4) surfaces were first obtained by growing an undoped homoepitaxial layer (~0.5μm) on a thermally cleaned GaAs(001) substrate. A detailed description of the apparatus and surface cleaning treatments for the GaAs(001)-(2x4) surface has been given in our previous papers.[12] The sample was then transferred via ultra-high vacuum (UHV) transfer modules to another UHV chamber, where surfaces of GaAs(001)-c(8x2) were obtained by heating the GaAs(001)-(2x4) surfaces above 600°C without As fluxes. The RHEED rocking curves were measured at temperatures higher than 600°C, and at a base pressure of $5 \times 10^{-11}$ Torr.

For RHEED rocking-curve measurements, the glancing angle of the incident electron beam (15 keV) was changed with intervals of ~0.04° using the extended beam rocking facility (Staib, EK-35-R and k-Space, kSA400). Integrated intensities of the 17 spots, (0 0), $(\pm\frac{1}{4}\ 0)$, $(\pm\frac{2}{4}\ 0)$, $(\pm\frac{3}{4}\ 0)$, $(\pm 1\ 0)$, $(\pm\frac{5}{4}\ 0)$, $(\pm\frac{6}{4}\ 0)$, $(\pm\frac{7}{4}\ 0)$, and $(\pm 2\ 0)$ for the [110] direction, and 5 spots, (0 0), $(0\ \pm 1)$, and $(0\ \pm 2)$ for the $[1\bar{1}0]$ direction, measured by a charge-coupled-device camera with a microcomputer system, were used in a structure analysis. Averaging of the intensities of symmetry-equivalent spots led to nine and three independent rocking curves for the [110] and $[1\bar{1}0]$ directions, respectively.

Intensities of RHEED were calculated by the multislice method proposed by Ichimiya.[14] 25 and 9 beams on the zeroth Laue zone were used in the calculation with the incident electron beam along the [110] and $[1\bar{1}0]$ directions, respectively: (0 0), $(\pm\frac{1}{4}\ 0)$, •••••, $(\pm\frac{11}{4}\ 0)$, and $(\pm 3\ 0)$ for the [110] direction, and (0 0), $(0\ \pm 1)$, $(0\ \pm 2)$, and $(0\ \pm 3)$ for the $[1\bar{1}0]$ direction. Fourier coefficients of the elastic scattering potential were obtained from the atomic scattering factors for free atoms calculated by Doyle and Turner.[15] A correction due to condensation was made to fit the positions of bulk Bragg peaks at large glancing angles. For instance, the resulting mean inner potential of bulk GaAs was 13.6 eV. The Debye-Waller parameters were taken to be 1.70 Å$^2$ and 1.47 Å$^2$ for Ga and As atoms in bulk layers,[16] while those for atoms in the surface bilayer were treated as fitting parameters. In order to quantify the agreement between the calculated rocking curves and the experimental ones, the R factor defined in ref. 11 was used. Details about the calculations were given in ref. 12.

When the GaAs(001)-(2x4) surface was heated above 450°C, the (2x4) reflections disappeared and 1/6-order spots became clearly visible along the $[1\bar{1}0]$ direction. Simultaneously, 1/2- and 1/3-order streaks begin to appear in the RHEED patterns obtained along the [110] direction. Our STM observations at room temperature have revealed that the surface has the (2x6)/(3x6) structure at this stage. As the temperature is increased above ~580°C the reflections associated with a c(8x2) reconstruction emerged and the (2x6)/(3x6) reflections disappeared at ~600°C. The c(8x2) structure is stable in the range of 600~640°C, beyond which the surface begins to roughen. Here, we note that the surface at 600~640°C has the c(8x2) periodicity, but not the (4x2) one: the 1/4-order spots lying on a semicircle, the zeroth-order Laue zone, are clearly observed in the RHEED patterns obtained along the [110] direction. On the other hand, spots are observed for the integral order reflections, but not for half-order ones along the $[1\bar{1}0]$ direction. Instead, spots are clearly visible on the 1/8-order Laue zone.

As the substrate temperature is decreased below 600°C, 1/6-order reflections appears in the RHEED pattern obtained along the $[1\bar{1}0]$ direction, and 1/2- and 1/3-order streaks begin to coexist with the 1/4-order spots in the [110]-RHEED patterns. Our STM observations have confirmed that the c(8x2) structure was partially preserved on the surface only when the samples were quenched from 600°C to room temperature. On the other hand, when the substrate temperature is gradually decreased below ~550°C, the c(8x2) structure completely disappeared. From these results, one may attribute the structural change to the adsorption of As molecules on the c(8x2) surface from residual gases. However, since the present experiments were performed in a good UHV condition of ~5x10$^{-11}$ Torr, we can rule out possible adsorption of As molecules. Thus, these results prompted us to study the atomic structure of the c(8x2) surface at temperatures higher than 600°C.

Figure 2 shows RHEED rocking curves measured from the GaAs(001)-c(8x2) surface at 610°C (solid curves). The shape of these curves are insensitive to the change in substrate temperature in the range of 600 to 640°C. Also shown in Fig. 2 are the calculated rocking curves from the optimized structure model shown in Fig. 3, which is basically the same as that shown in Fig. 1(e). The present analysis assumed that the surface relaxation extends no further than the third atomic layer. The R factor for the optimized model is 0.104, showing a good agreement between the experiment and calculation. On the other hand, the É$f$ model [Fig. 1(d)] gives a larger R factor of 0.154. We have also tested other structure models shown in Figs. 1(a)-1(h).[16] As shown in Fig. 4, most of structure models give R factors larger than 0.25, even after the structure optimization.

The structure parameters of the optimized model (Fig. 3) are listed in Table I with errors evaluated from the half width of the range where R factor is smaller than $1.1 \times R_{min}$. Comparing the atomic coordinates obtained by the present analysis and those in ref. 9, a good agreement is found between them. In addition, the site occupancies (ÉΔ) for Ga adatoms [Ga(1)] and surface Ga-dimer [Ga(2)] are 0.27±0.09 and 0.47±0.13, respectively, which are close to the corresponding values obtained by XRD analysis.[8,9] The reduced site occupancy of Ga(2) suggest that the atomic arrangement without Ga(2) atoms locally exists on the c(8x2) surface, but further studies are needed to examine the stability of such a local structure.

As seen in Table I, the atomic coordinates of Ga(1) and Ga(2) atoms have relatively large errors in the [110] direction. This is not due to a small data set in this direction, because other atoms, such as As(1), Ga(5) and Ga(6), have much smaller errors. On the other hand, when in-plane anisotropies in the Debye-Waller parameters were taken into account in the analysis, we found that the vibrational amplitudes of Ga(1) are 0.32 Å and 0.19 Å in the [110] and [1$\bar{1}$0] directions, respectively. Thus, the present result show that Ga(1) atoms are weakly bonded to the surface and are in uncertain positions in the [110] direction. This is consistent with the recent first-principles calculations showing that the existence of Ga(1) adatom is energetically unstable.[7]

The large errors in the Ga(2) position can be correlated with the following analysis. We have calculated RHEED intensities for the structure model in which Ga(2) atoms do not form dimers. This model arrived at an R factor of 0.106, for which the amount of Ga(2) atoms in a c(8x2) unit cell is 0.84 (ÉΔ=0.42). In addition, when the amount of Ga(2) atoms is in the range of 0.84~0.94, the model revealed clear minimum in an R factor of 0.10~0.11, irrespective of their atomic coordinate in the [110] direction. While such a random positioning of Ga(2) atoms results in the (4x1) periodicity in the outermost layer, the subsurface dimerization of Ga(6) produces the c(8x2) periodicity. From these results, we conclude that the formation of Ga(2) dimers is not necessarily favorable at a high temperature of 610°C [20]. Considering that the local structure without Ga(2) atoms can also exist on the GaAs(001)-c(8x2) surface at high temperatures, it seems reasonable to consider that thermally-activated Ga(2) atoms migrate along the [110] direction, changing their local atomic configurations.

In conclusion, we have studied the surface structure of GaAs(001)-c(8x2) at high temperatures. The c(8x2) structure obtained by heating the (2x6)/(3x6) surface above 600°C is stable at 600~640°C, but reversibly changes to the (2x6)/(3x6) surface as the temperature is decreased below 600°C. We performed RHEED rocking-curve analysis above 600°C and confirmed that the c(8x2) surface has the structure model proposed by Kumpf et al. The results suggest that surface Ga atoms are thermally activated at high temperatures and dynamically change their local atomic configurations without disturbing the c(8x2) symmetry.

We are indebted to Dr. T. Hanada for use of the RHEED intensity calculation program. This study was partly supported by NEDO.

**REFERENCES**

\* Author to whom correspondence should be addressed. Electronic mail: OHTAKE.Akihiro@nims.go.jp.

[1] See, for example, Q. -K. Xue, T. Hashizume, and T. Sakurai, Prog. Surf. Sci. **56**, 1 (1997).

[2] J. Cerdá, F. J. Palomares, and F. Soria, Phys. Rev. Lett. **75**, 665 (1995).

[3] J. E. Northrup and S. Froyen, Phys. Rev. Lett. **71**, 2276 (1993); Phys. Rev. B **50**, 2015 (1994).

[4] S. L. Skala, J. S. Hubacek, J. R. Tucker, and J. W. Lyding, S. T. Chou, and K. -Y. Cheng, Phys. Rev. B **48**, 9138 (1993).

[5] Q. Xue, T. Hashizume, J. M. Zhou, T. Sakata, T. Ohno, and T. Sakurai, Phys. Rev. Lett. **74**, 3177 (1995).

[6] S. H. Lee, W. Moritz, and M. Sheffler, Phys. Rev. Lett. **85**, 3890 (2000).

[7] D. Paget, Y. Garreau, M. Sauvage, P. Chiaradia, R. Pinchaux, and W. G. Schmidt, Phys. Rev. B **64**, 161305 (2001).

[8] C. Kumpf, L. D. Marks, D. Ellis, D. Smilgies, E. Landemark, M. Nielsen, R. Feidenhans'l, J. Zegenhagen, O. Bunk, J. H. Zeysing, Y. Su, and R. L. Johnson, Phys. Rev. Lett. **86**, 3586 (2001).

[9] C. Kumpf, D. Smilgies, E. Landemark, M. Nielsen, R. Feidenhans'l, O. Bunk, J. H. Zeysing, Y. Su, R. L. Johnson, L. Cao, J. Zegenhagen, B. O. Fimland, L. D. Marks, and D. Ellis, Phys. Rev. B **64**, 075307 (2001).

[10] S. Kohmoto and A. Ichimiya, Surf. Sci. **223**, 400 (1989); A. Ichimiya, S. Kohmoto, T. Fujii, and Y. Horio, Appl. Surf. Sci. **41/42**, 82 (1989).

[11] T. Hanada, S. Ino, and H. Daimon, Surf. Sci. **313**, 143 (1994); T. Hanada, H. Daimon, and S. Ino, Phys. Rev. B **51**, 13320 (1995).

[12] A. Ohtake et al., Phys. Rev. B **59**, 8032 (1999); Appl. Phys. Lett. **74**, 2975 (1999); Phys. Rev. B **60**, 8326 (1999); Phys. Rev. B **60**, 8713 (1999).

[13] J. M. McCoy et al., Surf. Sci. **261**, 29 (1992); Phys. Rev. B **48** 4721 (1993); Surf. Sci. **418**, 273 (1998).

[14] A. Ichimiya, Jpn. J. Appl. Phys., Part 1 **22**, 176 (1983); **24**, 1365 (1985).
7

**FIGURE CAPTIONS**

FIG. 1. Structure models for the GaAs(001)-c(8x2) surface.

FIG. 2. RHEED rocking curves (solid curves) measured from the GaAs(001)-c(8x2) surface at 610°C. The dashed curves are calculated from the optimized structure [Fig. 3].

FIG. 3. Optimized structure model for the GaAs(001)-c(8x2) surface.

FIG. 4. R factors for the various structure models for the GaAs(001)-c(8x2) surface.

TABLE I; Atomic coordinates of the optimized structure model for the GaAs(001)-c(8x2) surface. The atomic coordinates x and y are given as fraction of unit cells along the $[1\bar{1}0]$ and [110], directions in Fig. 3. The z coordinates refers to the [001] direction with magnitude equal to the bulk (001) spacing of 5.6538 Å, the origin of which is at the fifth atomic layer.

**FIG. 1. Ohtake et al.**

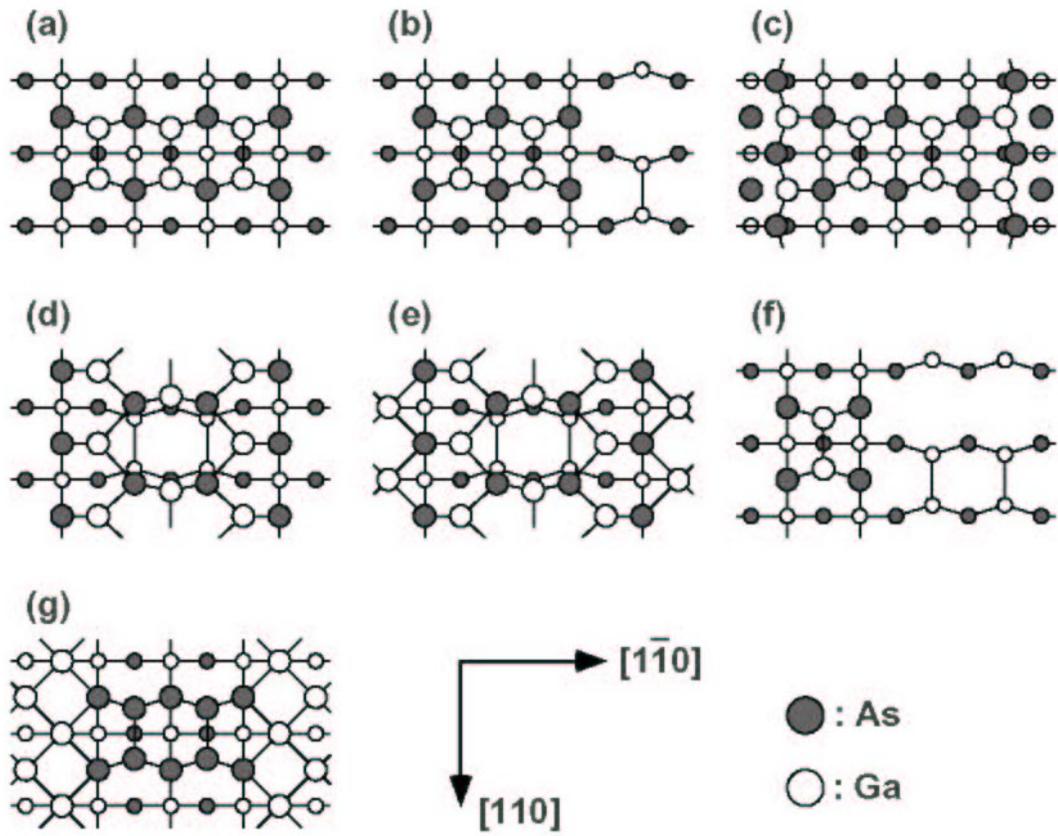

**FIG. 2.** Ohtake et al.

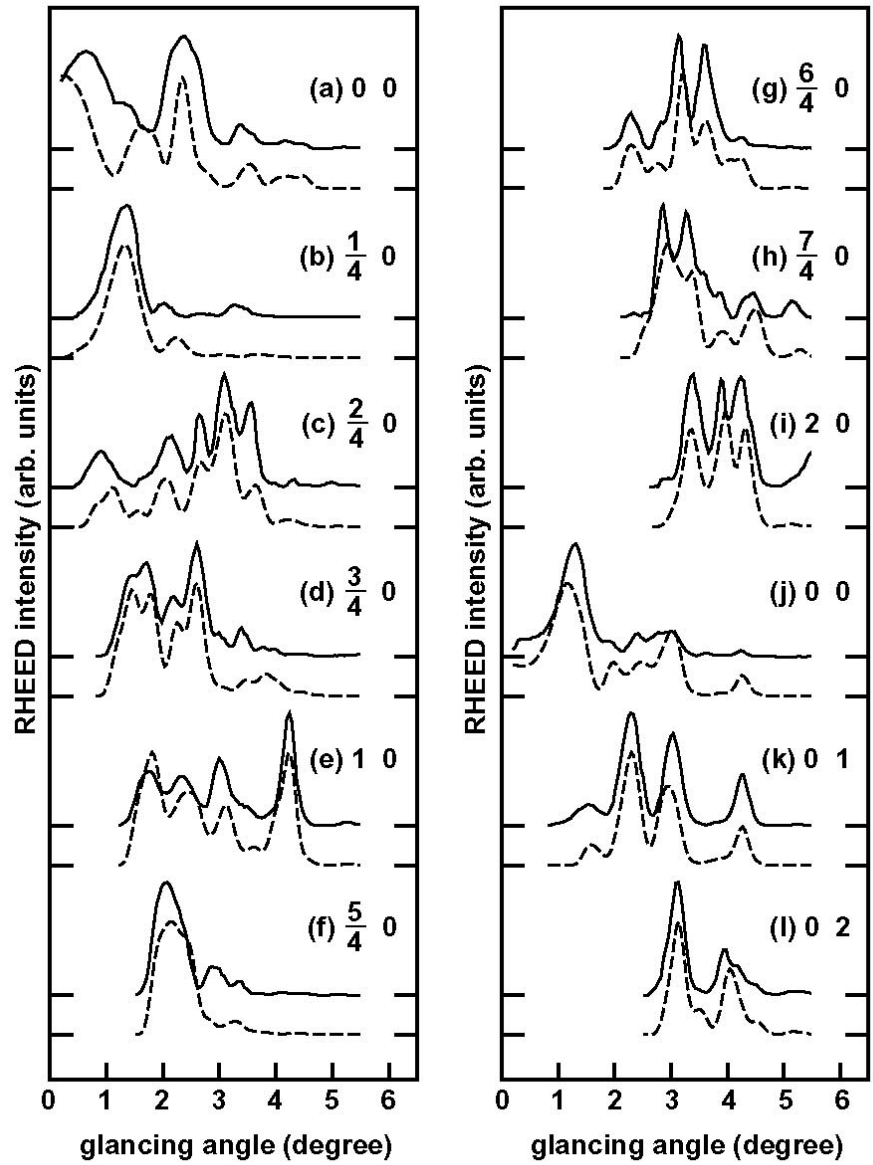

**FIG. 3. Ohtake et al.**

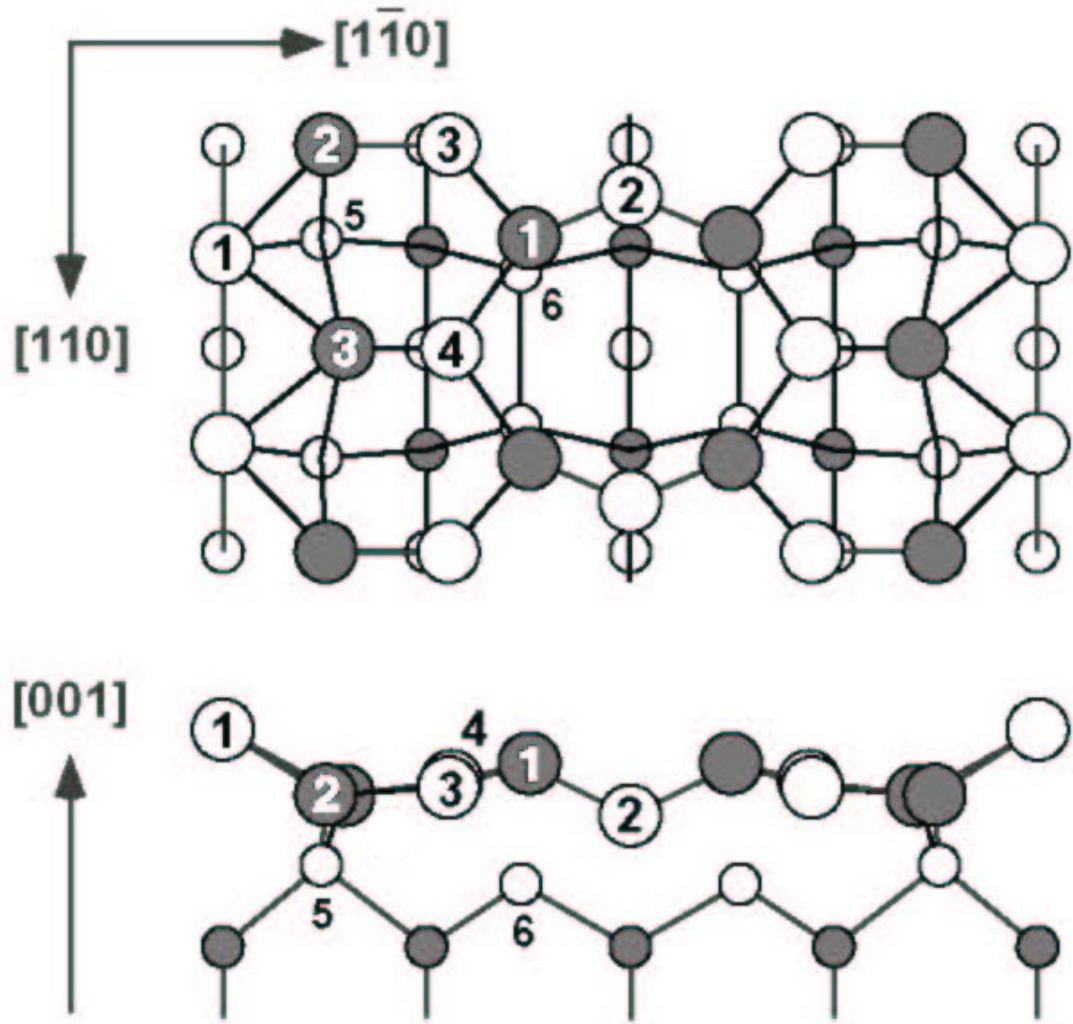

**FIG. 4. Ohtake et al.**

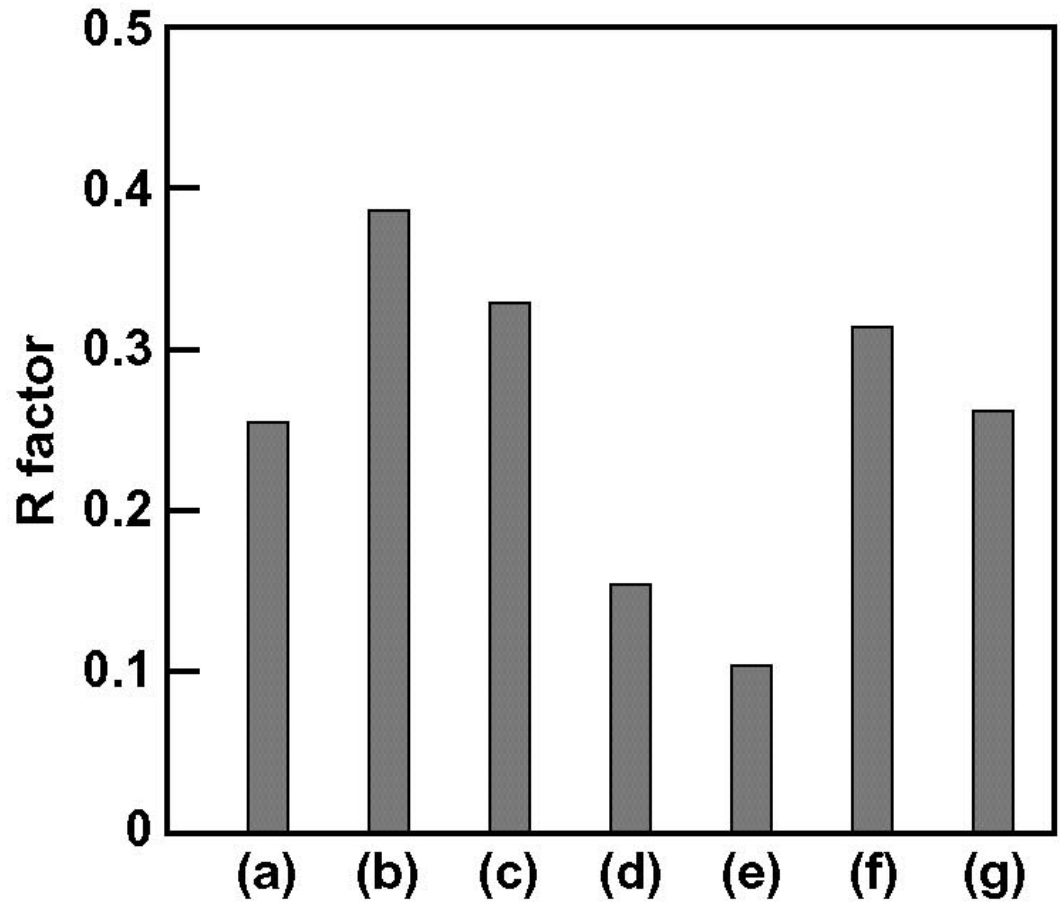

**TABLE I Ohtake et al.**

| Notations in Fig. ? | This study | | | | Kumpf et al. (ref. 9) | | | |
|---|---|---|---|---|---|---|---|---|
| | x | y | z | | x | y | z | |
| Ga(1) | 0.000 | 0.530±0.150 | 0.755±0.026 | 0.27±0.09 | 0.000 | 0.500 | 0.827 | 0.19 |
| As(1) | 1.501±0.020 | 0.457±0.030 | 0.638±0.016 | | 1.460 | 0.489 | 0.639 | |
| Ga(2) | 2.000 | 0.248±0.105 | 0.456±0.032 | 0.47±0.13 | 2.000 | 0.293 | 0.475 | 0.63 |
| Ga(3) | 1.108±0.040 | 0.000 | 0.554±0.021 | | 1.117 | 0.000 | 0.571 | |
| Ga(4) | 1.125±0.040 | 1.000 | 0.577±0.030 | | 1.130 | 1.000 | 0.579 | |
| As(2) | 0.505±0.045 | 0.000 | 0.528±0.023 | | 0.526 | 0.000 | 0.515 | |
| As(3) | 0.593±0.033 | 1.000 | 0.527±0.021 | | 0.539 | 1.000 | 0.509 | |
| Ga(5) | 0.483±0.030 | 0.463±0.032 | 0.283±0.012 | | 0.516 | 0.470 | 0.256 | |
| Ga(6) | 1.463±0.040 | 0.623±0.018 | 0.218±0.017 | | 1.474 | 0.670 | 0.212 | |